\documentclass[
 aip, apl,
 amsmath,amssymb,
 reprint,%
]{revtex4-1}

\usepackage{graphicx}
\usepackage{dcolumn}
\usepackage{bm}

\usepackage[utf8]{inputenc}
\usepackage[T1]{fontenc}
\usepackage{mathptmx}
\usepackage{etoolbox}
\usepackage{siunitx}
\usepackage{array}

\makeatletter
\def\@email#1#2{%
 \endgroup
 \patchcmd{\titleblock@produce}
  {\frontmatter@RRAPformat}
  {\frontmatter@RRAPformat{\produce@RRAP{*#1\href{mailto:#2}{#2}}}\frontmatter@RRAPformat}
  {}{}
}%
\makeatother
\begin{document}

\preprint{AIP/123-QED}

\title[Sample title]{Impact of Nitrogen and Oxygen Interstitials on Niobium SRF Cavity Performance
}
\author{H. Hu}
 \affiliation{Physics Department, The University of Chicago, Chicago, Illinois, 60637, USA}
 \affiliation{ 
Fermi National Accelerator Laboratory, Batavia, Illinois 60510, USA
}%

\author{Y.-K. Kim}
\affiliation{%
Physics Department, The University of Chicago, Chicago, Illinois, 60637, USA
}%
\affiliation{ 
Fermi National Accelerator Laboratory, Batavia, Illinois 60510, USA
}%
\author{D. Bafia}%
\affiliation{ 
Fermi National Accelerator Laboratory, Batavia, Illinois 60510, USA
}%
  \email{dbafia@fnal.gov}
\date{\today}

\begin{abstract}
Superconducting radiofrequency (SRF) cavities are the leading technology for highly efficient particle acceleration, and their performance can be significantly enhanced through the controlled introduction of interstitial impurities into bulk niobium. Nitrogen doping has demonstrated a substantial reduction in surface resistance, which improves the quality factor of the cavities. More recently, oxygen doping has emerged as a promising alternative, demonstrating comparable reductions in surface resistance. In this study, we combine cavity measurements on \SI{1.3}{GHz} niobium SRF cavities subjected to a range of nitrogen- and oxygen-based treatments with material characterizations performed on cavity cutouts processed under identical conditions. This approach allows us to quantitatively assess the contribution of each impurity to the reduction of surface resistance. We find that nitrogen is ten times more effective than oxygen in reducing surface resistance at \SI{16}{MV/m}. We also observe an additive effect of O and N impurities in reducing R$_\mathrm{T}$.  We discuss these results in the context of field dependent nonequilibrium superconductivity, gap enhancement, and hydrogen trapping mechanisms.

\end{abstract}

\maketitle

Extending the performance of superconducting radiofrequency (SRF) cavities is critical for enabling next-generation superconducting accelerators, with higher energies, reduced power losses, and lower cryogenic costs\cite{padamsee_rf_2008,isagawa_influence_1980}. In SRF cavities, electromagnetic fields build up within the cells; these fields accelerate charged particles along the cavity axis and, at the same time, induce surface currents that penetrate $\sim$\SI{100}{nm} into the superconducting surface, defining the RF layer\cite{isagawa_influence_1980, maxfield_superconducting_1965}. Over the past decade, significant improvements in performance have been made by modifying the composition of the RF layer through the introduction of interstitial impurities.

Surface treatments such as nitrogen doping, nitrogen infusion, and more recently, oxygen-based baking treatments, have been developed to enhance SRF performance. Nitrogen doping is one surface treatment which introduces uniform, dilute concentrations of N throughout the RF layer and yields high quality factors (Q$_0$) \cite{grassellino_nitrogen_2013}. In contrast, nitrogen infusion is a treatment which deposits a higher concentration of N confined to the first $\sim$\SI{20}{nm} of the surface, enabling higher accelerating gradients (E$_\mathrm{acc}$) rather than ultra-high Q$_0$\cite{grassellino_accelerating_2018}. While N doping has been successfully industrialized and implemented in full-scale cryomodules, N~infusion has proven more challenging to translate and apply outside of Fermilab \cite{galayda_lcls-ii_2018, raubenheimer_lcls-ii-he_2018, grassellino_unprecedented_2017}. More recently, oxygen-based surface treatments have recently emerged as a promising and reliable alternative \cite{bafia_role_2022,ito_influence_2021,lechner_rf_2021}. These treatments require only a single baking step with no subsequent electropolishing (EP), taking advantage of the naturally occurring \SI{5}{nm} Nb oxide; baking dissolves the oxide and drives inward O diffusion \cite{bafia_role_2022, veit_oxygen_2019,bafia_oxygen_2024}. Oxygen treatments can be categorized by temperature. Low-temperature baking at \SI{120}{\degreeCelsius} enhances accelerating gradients in a manner similar to N infusion, with O concentrated in the first $\sim$\SI{100}{nm}\cite{bafia_gradients_2019, bafia_role_2022, padamsee_rf_2009}. Medium-temperature (mid-T) baking (\SIrange{200}{350}{\degreeCelsius}) yields more uniform distributions of O and increases Q$_0$ akin to N doping \cite{posen_ultralow_2020, ito_influence_2021}. These treatments have also been applied to full-scale cryomodules and are under consideration for future major accelerators, including two-step low temperature baking for the International Linear Collider and mid-T baking for the Future Circular Collider and Shanghai High Repetition Rate XFEL and Extreme Light Facility (SHINE) \cite{katayama_high-qhigh-g_2022, bafia_gradients_2019, pan_high_2024}. Collectively, these results demonstrate that nitrogen- and oxygen-based approaches can be tailored for either high-gradient or high-Q$_0$ performance, motivating a direct comparison of their underlying mechanisms and the magnitudes of their effects.

\begin{table*}[t]
\caption{Summary of nitrogen- and oxygen-based surface treatments applied to single-cell \SI{1.3}{GHz} TESLA-shaped cavities and cutouts.}
\label{tab1}

\begin{tabular*}{\linewidth}{@{\extracolsep{\fill}} c c c}
\hline
\textbf{Cavity} & \multicolumn{2}{c}{\textbf{Treatment Steps}} \\
\hline\hline
TE1AES010 & EP Baseline &  \\
TE1AES010 & 120 $^\circ$C $\times$ 3 h \textit{in-situ} & + 120 $^\circ$C $\times$ 3 h (total 6 h) \textit{in-situ} \\
TE1PAV009 & 120 $^\circ$C $\times$ 48 h \textit{in-situ} &  \\
TE1AES017 & 200 $^\circ$C $\times$ 1 h \textit{in-situ} & + 200 $^\circ$C $\times$ 10 h (total 11 h) \textit{in-situ} \\
TE1AES021 & 200 $^\circ$C $\times$ 20 h \textit{in-situ} &  \\
TE1AES024 & 2/0+5 \textmu m N Doped &  \\
TE1RI003 & 3/60+10 \textmu m N Doped & + 8 \textmu m EP (3/60+18 \textmu m N Doped) \\
TE1PAV012 & N Infused (120 $^\circ$C $\times$ 48 h with 25 mTorr N$_2$ injection) & \\
TE1PAV008 & 350 $^\circ$C $\times$ 2.5 h \textit{in-situ} & oxidized in air\\
\hline
\end{tabular*}
\end{table*}

The performance of SRF cavities is limited by the RF surface resistance, which can be expressed as the sum of a temperature-dependent resistance (R$_\mathrm{T}$) and a temperature-independent residual resistance (R$_{\mathrm{res}})$. R$_\mathrm{T}$ arises from dissipation associated with thermally excited quasiparticles in the superconducting state and can be described using Mattis-Bardeen theory, an extension of BCS theory which accounts for an applied RF field \cite{mattis_theory_1958, bardeen_theory_1957}. Interstitial impurities such as nitrogen and oxygen influence R$_\mathrm{T}$ by modifying the mean free path in the RF layer, which can optimize the superfluid density and suppress quasiparticle losses. R$_\mathrm{T}$ can be expressed as: $\mathrm{R}_\mathrm{T}(\mathrm{E_{acc}},\mathrm{T})= A(\ell)\frac{f_0^2}{\mathrm{T}}e^{-\Delta/k_B\mathrm{T}}$
where $A(\ell)$ is a parameter dependent on the electronic mean free path, $f_0$ is the cavity resonant frequency, $\Delta$ is the superconducting energy gap and $k_B$ is Boltzmann's constant \cite{halbritter_surface_1974}. R$_\mathrm{res}$ can arise from an increase in the number of subgap quasiparticle states, lossy oxides, and metallic inclusions, and vortex oscillations \cite{kubo_field_2019, bafia_quantifying_2025}. 

Despite empirical evidence for the benefits of N and O interstitials in the RF layer, the underlying physical mechanisms are not yet fully understood. Theoretical models suggest that interstitials modify key superconducting parameters, including the mean free path, penetration depth, and superconducting gap \cite{gurevich_theory_2017, gurevich_reduction_2014}. A mismatch between a doped (dirty) surface and a clean bulk can increase the superheating field and delay vortex penetration under increasing RF fields\cite{ngampreutikorn_effect_2019,kubo_field_2019,gurevich_maximum_2015}. In the nonlinear Meissner screening theory, interstitials optimize the mean free path, increasing the superfluid density and reducing losses \cite{sauls_theory_2022}. At higher RF fields, nonlinear effects, including pair breaking and vortex penetration, can increase R$_\mathrm{T}$, contributing to field-dependent degradation of cavity performance \cite{sauls_theory_2022}. Nonequilibrium superconductivity is another mechanism in which interstitial impurities modify the quasiparticle distribution from thermal equilibrium by enhancing recombination or altering inelastic scattering rates under RF excitation\cite{martinello_field_2018, gurevich_reduction_2014}. Eliashberg theory provides a microscopic description of these nonequilibrium effects, in which interstitial impurities modify gap anisotropy and inelastic scattering under the strong-coupling regime \cite{eliashberg_film_1970}.

Another proposed mechanism for enhancing SRF cavity performance is that interstitial impurities act as hydrogen traps \cite{ford_suppression_2013, isagawa_influence_1980, pfeiffer_trapping_1976}. Hydrogen is highly diffusive in Nb, making it particularly susceptible to hydrogen uptake. Additionally, many standard cavity processing techniques rely on acids, which may introduce additional H \cite{knobloch_q_2003, barkov_precipitation_2013}. When Nb is cooled below \SI{150}{K}, interstitial H may precipitate as niobium nanohydrides, which become superconducting through the proximity effect, thus degrading superconducting properties and increasing losses \cite{romanenko_proximity_2013, knobloch_q_2003}. The formation of these nanohydrides has been observed with cryogenic transmission electron microscopy and cryogenic atomic force microscopy\cite{trenikhina_nanostructural_2015, sung_direct_2024}. Introducing dilute concentrations of impurities, such as N, O, C, and Ti, can distort the bcc Nb lattice and allow the interstitials to act as effective traps for H\cite{koufalis_effects_2017, ford_first-principles_2013, dhakal_effect_2013}. First-principle calculations confirm it is more energetically favorable for interstitial H to bind to N or O than to Nb\cite{ford_first-principles_2013}.

In this work, we conduct a quantitative analysis of the effect of nitrogen and oxygen impurities on R$_\mathrm{T}$ by correlating cavity measurements to materials characterizations from time-of-flight secondary ion mass spectroscopy (ToF-SIMS). Our results indicate that nitrogen is up to ten times more effective than oxygen at \SI{16}{MV/m} at reducing R$_\mathrm{T}$. We interpret these findings in the context of previous experimental studies, Mattis-Bardeen theory, and theoretical models.

Single-cell \SI{1.3}{GHz} TESLA-shaped cavities fabricated from high-purity niobium were used to investigate the effects of N and O interstitials on RF performance. Cavity cutouts, \SI{1}{cm} in diameter, from TE1AES008 were prepared using the same surface treatment procedures as the cavities. Initial surface preparation followed a standard baseline procedure of a bulk 120~\textmu m EP removal, followed by \SI{800}{\degreeCelsius} degassing for 3~h in an ultra-high vacuum (UHV) furnace, 40~\textmu m EP removal, and high-pressure rinsing to clean the surface \cite{padamsee_rf_2008}. Subsequently, the treatments detailed in Table~\ref{tab1} were applied. All but one of the \textit{in-situ} baked cavities were assembled prior to baking and maintained vacuum through testing, whereas N doped, N infused, and EP cavities were assembled post-treatment. TE1PAV008 (data from \textcite{posen_ultralow_2020}) was \textit{in-situ} baked, vented with nitrogen, disassembled, exposed to cleanroom air for 10~min, and reassembled. After assembly, the vacuum level within each cavity was maintained below $< 1\times10^{-5}$~Torr.

To quantify the concentration of N and O impurities introduced into the cutouts after each processing technique outlined in Table~\ref{tab1}, we used ToF-SIMS to obtain depth profiles of the impurities present in the Nb lattice. Implanted N and O standards were measured in parallel to calibrate the relative ToF-SIMS intensity to absolute concentration, allowing a direct comparison between O and N impurities\cite{budrevich_metrology_1998}. Specifics on ToF-SIMS measurement are described in the supplementary material.

RF performance of each single-cell Nb cavity was evaluated at the Fermilab Vertical Test Stand (VTS) in continuous-wave operation using power balance measurements\cite{melnychuk_error_2014}. All cavities underwent a fast cooldown protocol to minimize magnetic flux trapping\cite{romanenko_dependence_2014, posen_efficient_2016}. Q$_0$ vs. E$_\mathrm{acc}$ curves were measured at both \SI{2}{K} and $<$\SI{1.5}{K} to decompose the surface resistance into R$_\mathrm{T}$ and R$_\mathrm{res}$\cite{martinello_effect_2016}. The temperature-dependent component at \SI{2}{K} can be obtained from R$_\mathrm{T}(\SI{2}{K})~=~$G$/$Q$_0$(\SI{2}{K}) $-$ G$/$Q$_0$($<$\SI{1.5}{K}), where G~=~270~$\Omega$ is a geometry factor that is only dependent on cavity shape. This study primarily investigates the evolution in R$_\mathrm{T}$ as our main interest is the effect of bulk impurities on superconducting properties.

\begin{figure}
\includegraphics[width=\columnwidth]{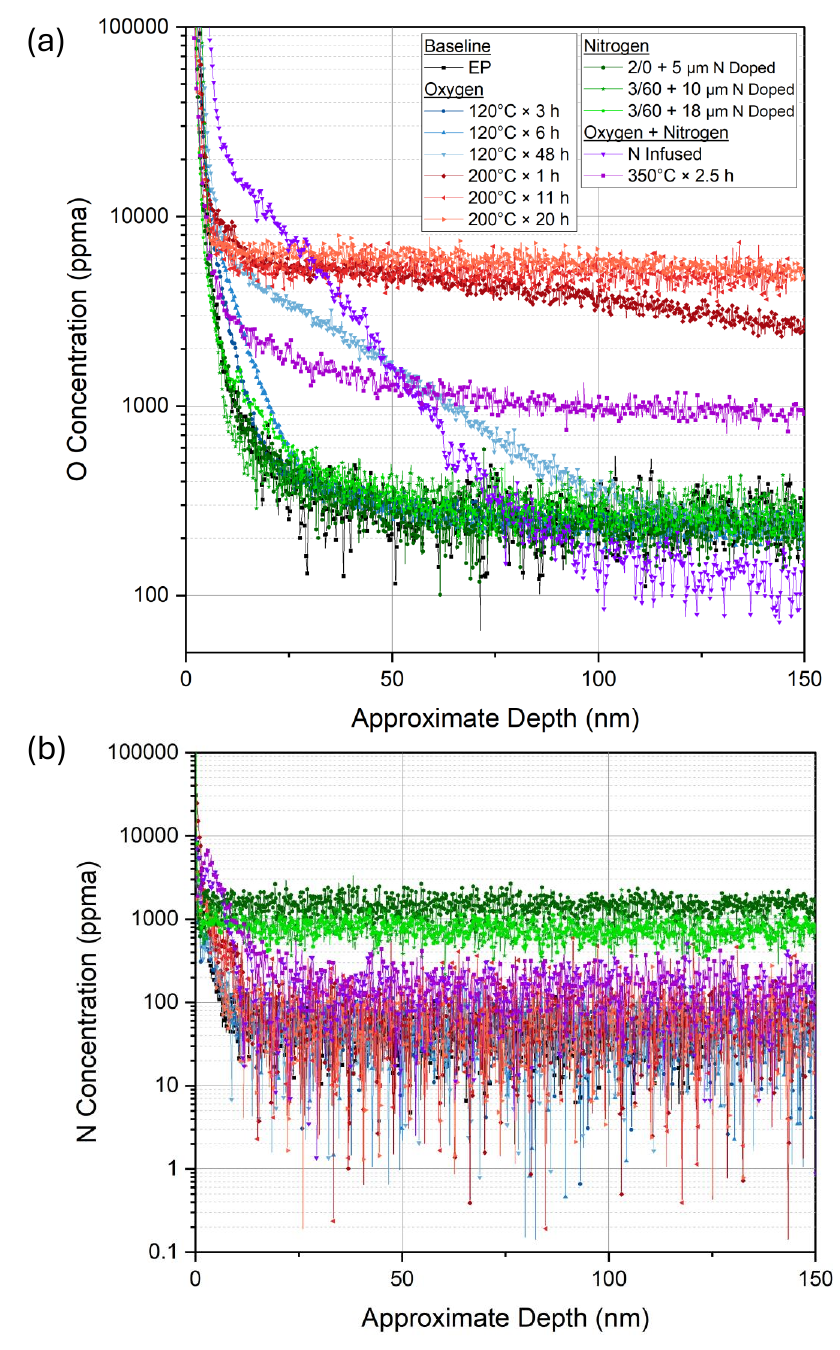}
\caption{\label{fig:f1}  ToF-SIMS depth profile of (a) oxygen and (b) nitrogen concentrations in Nb cavity cutouts after various surface treatments.}
\end{figure}

Figure~\ref{fig:f1} shows ToF-SIMS depth profile of O and N in Nb cavity cutouts for each treatment. Concentrations are presented in the dimensionless unit of parts per million atomic (ppma), which represents the number of impurity atoms per $10^6$ total atoms. The EP baseline exhibits a clean surface with no detectable O or N impurities. Baking at \SI{120}{\degreeCelsius} gradually dissolves oxygen into the bulk with a diffusion length of \SIrange{20}{100}{nm}, and baking at \SI{200}{\degreeCelsius} leads to a more uniform oxygen distribution within the RF layer \cite{bafia_role_2022}. No N was detected in the EP, 120$^\circ$C, or 200$^\circ$C baked samples above background levels. N doped samples show uniform and dilute concentrations of $\sim$10$^3$~ppma of N extending well beyond the RF layer, with negligible O concentration. The N infused sample contains N in the first $\sim$20~nm and O comparable to that of a 120$^\circ$C$\times$48~h bake. A mid-T baked (350$^\circ$C$\times$2.5~h) sample displays  N in the first $\sim$20~nm and a uniform concentration of O. For both N infused and mid-T baked samples, we detected nitrogen within the bulk above background levels. These treatments enable consideration of treatments that introduce both O and N. It is unclear what the source of N is in mid-T baking, but these findings mirror the nitrogen interstitials observed in \textcite{posen_ultralow_2020} of mid-T baked samples. Additional ToF-SIMS depth profiles of C, OH, H, and NbH are provided in the supplementary material along with quantification of trace contaminants that are only present in the first 5~nm of the surface.

\begin{figure}
\includegraphics[width=0.93\columnwidth]{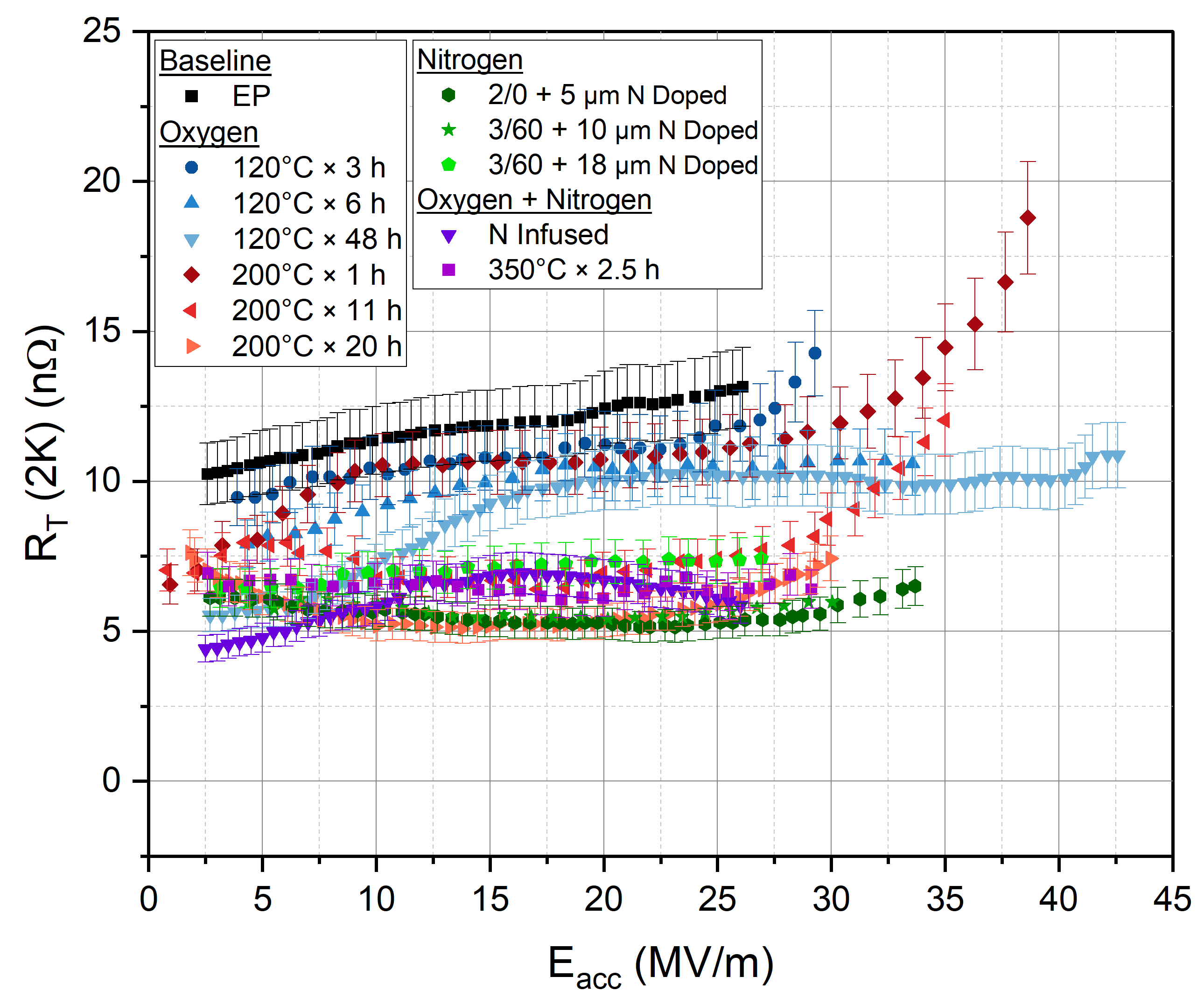}
\caption{\label{fig:f3} R$_\mathrm{T}$ vs. E$_\mathrm{acc}$ for single-cell 1.3~GHz SRF cavities processed with various treatment recipes.}
\end{figure}

Figure \ref{fig:f3} plots the corresponding RF measurements of R$_\mathrm{T}$ as a function of E$_{\mathrm{acc}}$ for each treatment. The reported uncertainty is 10\%, an upper limit on the error \cite{melnychuk_error_2014}. Longer 120$^\circ$C bakes yield a gradual increase in maximum E$_{\mathrm{acc}}$, consistent with previous studies \cite{romanenko_effect_2013, bafia_role_2022}. Both 200$^\circ$C \textit{in-situ} baking and N doping substantially reduce R$_\mathrm{T}$. Notably, the performances of the 200$^\circ$C$\times$20~h baked and  350$^\circ$C$\times$2.5~h baked cavities are comparable to that of both the 2/0+5~\textmu m and 3/60+10~\textmu m N doped treatments, indicating that O and N can play analogous roles in mitigating BCS-related losses. No significant change in R$_\mathrm{T}$ is expected if instead of \textit{in-situ} baking, the cavities were furnace baked and re-exposured to air to allow for oxide formation. The supplementary material presents an additional study in which a partially dissolved oxide is fully regrown with no changes in R$_\mathrm{T}$. 

\begin{figure}
\includegraphics[width=0.97\columnwidth]{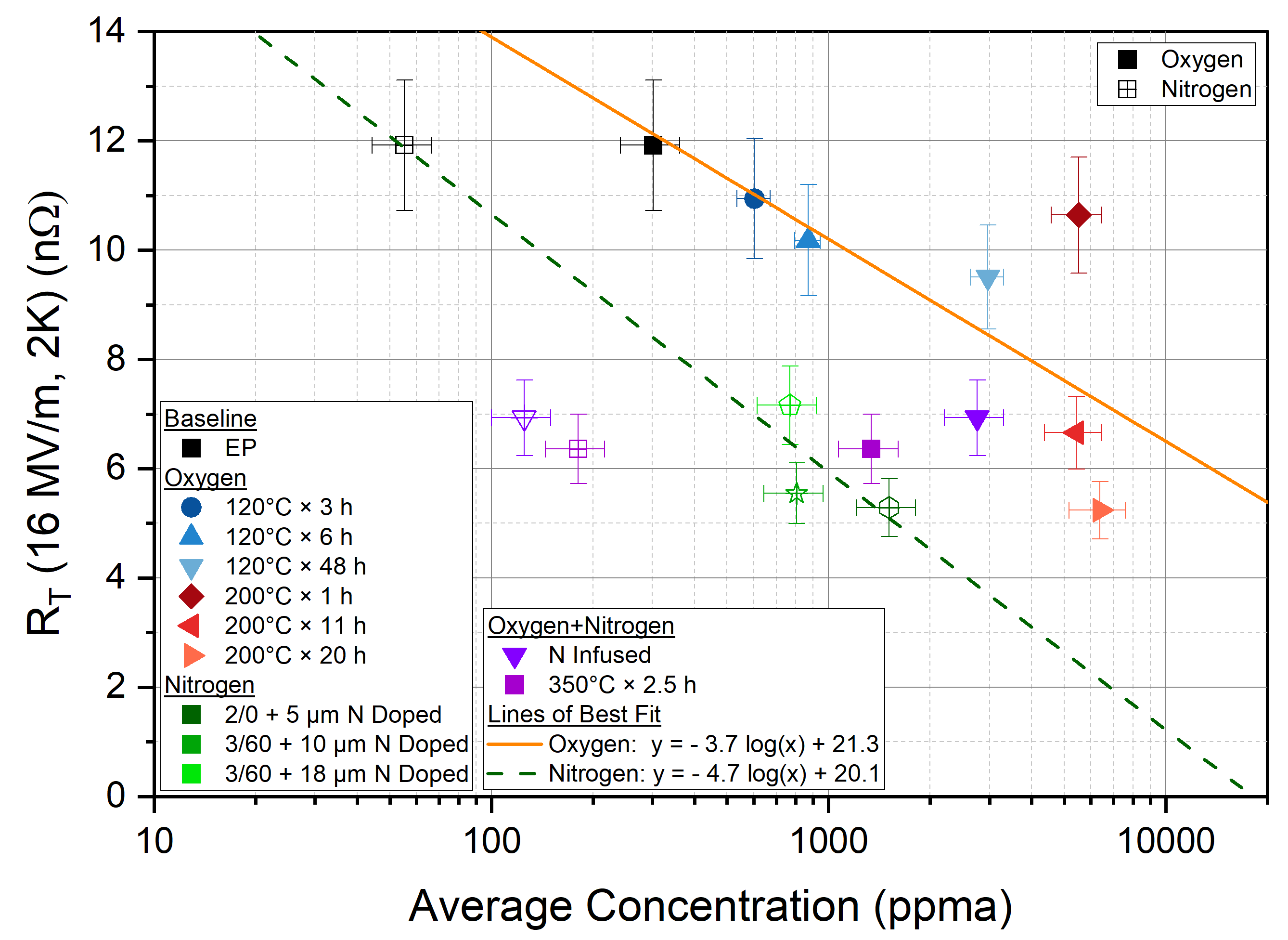}
\caption{\label{fig:f4} Correlation of R$_\mathrm{T}$ (16~MV/m, 2~K) with average O and N impurity concentration in the RF layer, excluding the first \SI{15}{nm} to avoid surface oxide counts.}
\end{figure}

A direct correlation between cavity RF performance and impurity concentration is shown in Fig.~\ref{fig:f4}. We extracted R$_\mathrm{T}$ at 16~MV/m for each of the treatments from Fig.~\ref{fig:f3} and averaged impurity concentrations from Fig.~\ref{fig:f1} over the RF layer, excluding the top \SI{15}{nm} to remove surface oxide contributions. R$_\mathrm{T}$ for both O- and N-based treatments exhibit approximately logarithmic scaling with concentration, where higher interstitial concentrations systematically lower R$_\mathrm{T}$. The lines of best fit indicate that nitrogen provides a stronger suppression per unit concentration than oxygen. At 16~MV/m, N doped cavities display R$_\mathrm{T}$ values similar to those of 200$^\circ$C$\times$20~h but with an order of magnitude less interstitial content. Thus, either species can drive comparable performance improvements when present at sufficient levels. It is surprising that there is such a large difference in the effects of N and O, given that both N and O can only trap one H per atom \cite{morkel_nitrogen-hydrogen_1978, richter_diffusion_1978}. It may be that hydrogen trapping is not the sole mechanism which drives differences between N and O.

N infusion and mid-T baking provide hybrid cases. In Fig.~\ref{fig:f4}, N infusion represented twice: once by its O concentration, comparable to a 120$^\circ$C$\times$48~h bake, and once by its N concentration. Relative to a 120$^\circ$C$\times$48~h bake, the N infused cavity shows an additional reduction in R$_\mathrm{T}$ of 2.5~n$\Omega$, attributable to the small addition of N. This suggests that the combined presence of O and N interstitials can have an additive effect in lowering R$_\mathrm{T}$. A similar additive effect can also be observed for mid-T baking.
 
\begin{figure}
\includegraphics[width=\columnwidth]{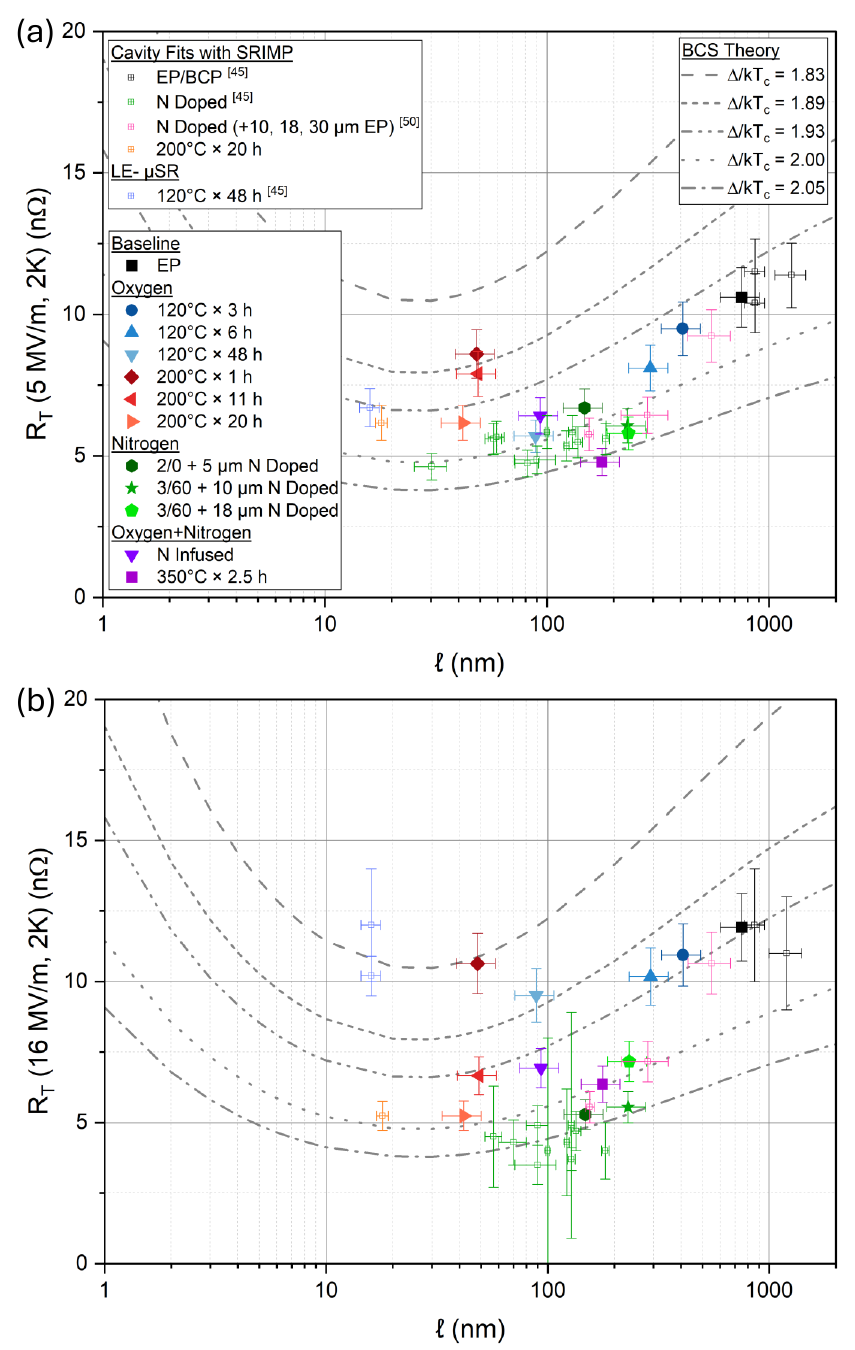}
\caption{\label{fig:f5} Mean free path ($\ell$) extracted from impurity concentrations plotted against (a) R$_\mathrm{T}$ (5 MV/m, 2K) and (b) R$_\mathrm{T}$ (16 MV/m, 2K) compared to theoretical BCS calculations and literature results of $\ell$ extracted using SRIMP and from LE-\textmu SR.}
\end{figure}

To contextualize these results, we compare them with BCS theory and previous experimental studies. Mean free path analysis (Fig.~\ref{fig:f5}) provides a unifying framework for O and N.  To take into consideration the evolution of R$_\mathrm{T}$ with field, we extracted R$_\mathrm{T}$ at both low field (5~MV/m) and medium field (16~MV/m). The supplemental material contains analysis at high fields (25~MV/m). Average impurity concentrations were converted into mean free path ($\ell$) using $\ell = 1/(c \pi r^2)$, where $c$ is the ion concentration in ions/cm$^3$and $r=1.46$~\AA~is the atomic radius of Nb. Ion concentration can be obtained from ppma using the Nb atomic density of $5.56\times10^{22}$ atoms/cm$^3$. This method of conversion relies on a simplified scattering assumption which estimates $\ell$ from the number of collisions expected within a scattering cross section. We take $c$ to be the sum of the concentrations of O and N, ignoring contributions from other impurities. This method provides a comparative estimate of $\ell$ across treatments. For reference, the theoretical BCS curves for various $\Delta/kT_c$ were calculated using SRIMP with fixed parameters: $T_c = 9.25$~K, $f_0 = 1.3$~GHz, $\lambda_L = 39$~nm, and $\xi = 38$~nm \cite{halbritter_fortran-program_1970}. Our results are also compared with experimentally determined $\ell$ values from \textcite{martinello_effect_2016} and \textcite{bafia_exploring_2021}, obtained through low-energy muon spin rotation \mbox{(LE-\textmu SR)} measurements and cavity $\ell$ fit and extracted with SRIMP \cite{halbritter_fortran-program_1970, martinello_effect_2016}. We observe good agreement between extracted $\ell$ for EP and N doped cavities across three independent measurement techniques: ToF-SIMS, \mbox{LE-\textmu SR}, and cavity fits with RF measurements using SRIMP. 

Overall, the data at 16 MV/m ((Fig.~\ref{fig:f5}(b)) highlight two distinct regimes. Treatments that confine impurities to the near-surface, such as baking at \SI{120}{\degreeCelsius} and N infusion, primarily reduce $\ell$ without significantly altering the superconducting gap. This results in significant improvement in E$_{\mathrm{acc}}$ but only limited improvement in Q$_0$. It is likely that these treatments form a dirty---clean superconducting bilayer, which can modify the local supercurrent density to act as an additional barrier against vortex penetration \cite{gurevich_theory_2017, kubo_multilayer_2016, lechner_oxide_2024}. In contrast, N doping reduces $\ell$ and simultaneously increases $\Delta$, resulting in a high Q$_0$ performance. Similarly, longer durations of baking at \SI{200}{\degreeCelsius} progressively lower R$_\mathrm{T}$ by enhancing $\Delta$, even when the average impurity concentration changes minimally. This could be attributed to a homogenization of $\Delta$ as observed in point-contact tunneling spectroscopy (PCTS) studies or to changes in the effective penetration depth \cite{groll_insight_2018,mcfadden_depth_2023}. These results demonstrate that both O and N reduce R$_\mathrm{T}$ through a combination of mean free path optimization and gap enhancement, with gap enhancement observed only in treatments exhibiting sufficient uniformity in the impurity depth profile throughout the RF layer. A more uniform impurity profile likely ensures that the entire RF penetration layer remains within the optimal scattering regime and enables a more homogeneous superconducting state. 

The observed gap enhancements are significantly less pronounced at 5 MV/m, indicating a field dependence consistent with nonequilibrium superconducting effects. Recent studies have shown that in fully gapped superconductors such as niobium, quasiparticle relaxation is dominated by electron–phonon scattering, which reduces the minimum frequency required for nonequilibrium behavior from the 15~GHz predicted by Eliashberg theory to $\sim$1.2~GHz \cite{eliashberg_film_1970, dobrovolskiy_moving_2020}. The cavities investigated in this work operate at 1.3~GHz, placing them within this nonequilibrium regime. This interpretation is further supported by the reported frequency dependence of the anti-Q slope in N doped cavities \cite{martinello_field_2018}. A corresponding systematic study of frequency dependence in O doped cavities would be required to determine whether the observed reduction in R$_\mathrm{T}$ can likewise be attributed to nonequilibrium superconductivity. 

With regard to the hydrogen trapping hypothesis, we observe that N doping increases the average superconducting gap at approximately ten times lower concentration than oxygen. This suggests that N has a stronger effect per atom, potentially due to more effective hydrogen trapping. N has been shown to suppress the formation of niobium nanohydrides \cite{sung_direct_2024}. Coincident with this, other studies have demonstrated that N doping reduces the smearing of the quasiparticle density of states compared to EP \cite{bafia_signatures_2025, groll_insight_2018}. These observations linking H minimization and reduced smearing are consistent with theoretical predictions \cite{kubo_effects_2022, herman_microwave_2021, zarea_effects_2023}. 

However, first-principle calculations by \textcite{ford_first-principles_2013} indicate only a minor difference in binding energies for \mbox{H-O}~($-$7.02~eV) and \mbox{H-N}~($-$7.39~eV), which insufficient to explained the observed order-of-magnitude difference. One possible explanation is that N doped surfaces form fewer vacancies and less defective suboxides than O doped surfaces, thereby enhancing their superconducting properties \cite{yang_xps_2018, wenskat_vacancy_2022}. X-ray photoelectron spectroscopy and transmission electron microscopy studies have also indicated that N doped surfaces form a more stoichiometric Nb$_2$O$_5$ layer with thinner metallic suboxide layers \cite{lechner_electron_2020, fang_understanding_2023}. Additional mechanisms may include the higher concentration of interstitial O inducing greater elastic strain on the niobium lattice than the lower concentrations of interstitial N; increases in strain in Nb have been shown to decrease T$_\mathrm{c}$ \cite{antoine_influence_2019, gardener_pressure_1966}. Moreover, interstitial oxygen suppresses T$_\mathrm{c}$ by $\sim0.9$~K per atomic \% O\cite{desorbo_effect_1963}.

In this work, we systematically investigated the effects of O and N impurities on the performance of niobium SRF cavities and their superconducting properties. By correlating material characterizations with RF performance, we demonstrated that O and N independently modify the superconducting properties of the RF layer. N doping achieves comparable R$_\mathrm{T}$ reduction at $\sim10\times$ lower impurity concentration than O-based treatments. The reduction in R$_\mathrm{T}$ may be attributable to gap enhancement and homogenization, as the presence of impurities may eliminate defects in the Nb lattice and act as traps for hydrogen. A field dependence in R$_\mathrm{T}$ is also indicative of nonequilibrium superconductivity, although further studies are necessary to confirm if this is also applicable to oxygen. In comparison, the effect of oxygen is more gradual as it must overcome losses introduced by increased vacancies in the surface oxide. This gradual effect, coupled with the simpler processing requirements for O-based treatments, highlights the potential of oxygen diffusion for finer tuning of the RF surface in future treatments. Additionally, the additive effect of O and N impurities in reducing R$_\mathrm{T}$ for N infusion indicates potential for combined O and N impurity tailoring. Further theoretical and experimental work is needed to fully understand the distinct roles of oxygen and nitrogen at the atomic scale. 

\section*{Supplementary Material}
See the supplementary material for details on ToF-SIMS analysis, additional ToF-SIMS data on surface contaminants and hydrogen trapping, high-field behavior of R$_\mathrm{T}$ at 25~MV/m, and a study demonstrating that partial vs fully grown surface oxide has no effect on R$_\mathrm{T}$.

\section*{ACKNOWLEDGEMENTS}
The authors would like to thank Alexandr Netepenko, Adam Clairmont, Akshay Murthy, Damon Bice, Tim Ring,  and Davida Smith for their support in sample and cavity preparation, data acquisition, and data analysis. This manuscript has been authored by the FermiForward Discovery Group, LLC under contract No. 89243024CSC000002 with the U.S. Department of Energy, Office of Science, Office of High Energy Physics. This work was supported by the University of Chicago. 
\nocite{*}
\bibliography{main}

\end{document}